\documentclass[aps,pre,preprint,groupedaddress,showpacs]{revtex4-1}
\pdfoutput=1
\usepackage{mathrsfs}
\usepackage{amssymb}
\usepackage{amsmath}
\usepackage{graphicx}
\newcommand{\beq}{\begin{equation}}
\newcommand{\eeq}{\end{equation}}
\newcommand{\vep}{\varepsilon}
\newcommand{\sig}{\sigma}
\newcommand{\lan}{\langle}
\newcommand{\ran}{\rangle}

\usepackage{color}

\begin{document}
\title{How to control Spin-Seebeck current in a metal-quantum dot-magnetic insulator junction}

\author{Lei Gu}
\affiliation{School of Physics and Wuhan National High Magnetic field center, Huazhong University of Science and Technology, Wuhan 430074, P. R. China}
\author{Hua-Hua Fu}
\email{hhfu@hust.edu.cn}
\affiliation{School of Physics and Wuhan National High Magnetic field center, Huazhong University of Science and Technology, Wuhan 430074, P. R. China}
\affiliation{Department of Physics and Astronomy, University of California, Irvine, California 92697-4575, USA}
\author{Ruqian Wu}
\email{wur@uci.edu}
\affiliation{Department of Physics and Astronomy, University of California, Irvine, California 92697-4575, USA}

\vspace{10pt}

\begin{abstract}
The control of the spin-Seebeck current is still a challenging task for the development of spin caloritronic devices. Here, we construct a spin-Seebeck device by inserting a strongly correlated quantum dot (QD) between the metal lead and magnetic insulator. Using the slave-particle approach and non-crossing approximation, we find that the spin-Seebeck effect increases significantly when the energy level of the QD locates near the Fermi level of the metal lead due to the enhancement of spin flipping and occurrences of quantum resonance. Since this can be easily realized by applying a gate voltage in experiments, the spin-Seebeck device proposed here can also work as a thermovoltaic transistor. Moreover, the optimal correlation strength and the energy level position of the QD are discussed to maximize the spin-Seebeck current as required for applications in controllable spin caloritronic devices.
\end{abstract}

\maketitle

\section{Introduction}

One of the most interesting topics in spin caloritronics~\cite{Goennen,Bauer} is the spin-Seebeck effect (SSE)~\cite{Uchida1}, in which pure spin current can be induced by a temperature gradient in materials or heterojunctions. SSE has been observed in magnetic metals~\cite{Uchida1}, semiconductors~\cite{Jaworski,Breton} and insulators~\cite{Uchida2,Uchida3,Kikkawa,Qu,Wu}, and various devices based on SSE have been proposed. Tremendous research interest has been inspired to study the fundamental mechanisms of SSE, and to explore new spin-Seebeck materials. Recent theoretical studies showed that spin-Seebeck rectification and negative differential SSE can exist in junctions of a metal and magnetic insulator (MI) if the electronic density of states (DOS) in the metal lead strongly fluctuates~\cite{Ren}. Moreover, the spin-Seebeck diode effect was proposed in either a spin valve nanopillar made of two permalloy circular disks ~\cite{Borlenghi}, or in a two-dimensional junction of functionalized materials~\cite{Fu}; both can produce unidirectional spin-Seebeck currents for spintronic applications.

MIs~\cite{Uchida2,Uchida3} are of particular interest since they can conduct pure spin currents without a motion of charge carriers. In a metal-MI junction, electrons in metal flip their spin direction when they absorb magnetization excitations (magnons in MI). When a temperature gradient is introduced, magnons can be driven away from their equilibrium state at the interface of MIs. As a sequence, a net spin current (i.e., the spin-Seebeck current) can be generated in metal by the flows of magnons in MI. Clearly, the control of electron-magnon coupling across the metal/MI interface is an effective route to optimize the spin-Seebeck current. In a quantum-dot (QD) or a junction, considering that the exclusion principle forbids double occupation of two electrons in the same quantum state, the spin flipping process is more likely to occur when a nearest energy level is either empty or occupied by an electron with the opposite spin direction. It is natural to expect that the SSE and spin-Seebeck current can be remarkably enhanced by introducing a QD into SSE devices.

\begin{figure}
\centering
\includegraphics[width=0.8\textwidth]{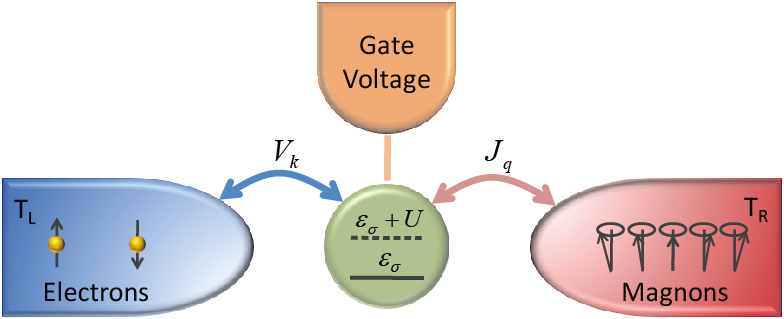}\\
\caption{Schematic of a strongly correlated quantum dot attached to the left metal lead and the right magnetic insulator with coupling $\Gamma_{k}$ and $J_{q}$, respectively. Here, \emph{U} is the Coulomb correlation, and $\vep_\sig$ denotes the dot level, which can be tuned by the top gate voltage. It is assumed that $J_{q}$ is much smaller than $\Gamma_{k}$ so that we approximate the spin current to the lowest order in $J_q$ and neglect the self-energy due to interaction with magnons.}\label{Figure_1}
\end{figure}

Here, we construct a spin-Seebeck device which has a QD inserted between the MI and metal lead, as sketched in Fig.~\ref{Figure_1}. It is known that the Coulomb blockade effect usually exists in QDs and most QD systems may exhibit rich thermoelectric behaviors. For example, the thermoelectric figure of merit of QD systems can be largely enhanced by increasing the Coulomb blockade effect~\cite{Xie}. In QD-based devices, beyond the studies of the thermopower of QDs~\cite{Scheibner,Costi,Andergassen,Ye,Taylor,Rejec,Weymann,Wojcik,Misiorny}, spin-dependent thermopower can be generated when QDs are coupled to magnetic leads~\cite{Rejec,Weymann,Wojcik} or are magnetized~\cite{Misiorny}. Obviously, the Coulomb interaction can control the rate of spin flipping in our device and is an effective way to optimize the spin-Seebeck current in QD systems or junctions. Indeed, our calculations show that the spin-Seebeck current can be significantly tuned by manipulating the energy levels of the QD, particularly when these energy levels are slightly below the Fermi energy of the metal leads. Since the energy levels in QD can be controlled by a top gate, the spin-Seebeck device proposed here can also be used as a thermovoltaic transistor.

The remainder of this paper is organized as follows. In Sec~\ref{sec:mm}, we present details of the model Hamiltonian and results of transport studies based on the non-crossing approximation (NCA) method~\cite{Bickers,Wingreen,Hettler} in terms of slave bosons~\cite{Barnes,Coleman,Read}. Results for two extreme cases, $U=0$ and $U=\infty$, are discussed in Sec~\ref{sec:nc} to see how strong correlation affects the SSE. In Section~\ref{sec:fc}, we generalize the NCA to situations with finite-$U$ and address the joint effect of the correlation and bias on the spin transport properties. At the end, we present the derivations of formulations in the Appendixes.

\section{Model and Method}
\label{sec:mm}
Without MI, the QD-metal system can be described by the Anderson impurity model. Using the slave-boson representation $c_{d\sig}^\dagger=f_\sig^\dagger b+\epsilon_{\sig\sig'} f_{\sig'}d^\dagger$ for electrons in QD, the Hamiltonian can be expressed as
\begin{align}
H_{\mathrm{imp}}&=\sum_{k,\sig}(\vep_k-\mu_\sig) c^\dagger_{k\sig}c_{k\sig}+\sum_{\sig}\vep_\sig f^\dagger_\sig f_\sig
+U d^\dagger d\nonumber\\
&+\sum_{k,\sig}(V_k c^\dagger_{k\sig}f_\sig b^\dagger+V_k^* f^\dagger_\sig c_{k\sig} b)
+\sum_{k,\sig\neq\sig'}(V_k c^\dagger_{k\sig}f^\dagger_{\sig'} d+V^*_k f_{\sig'} c_{k\sig} d^\dagger).
\end{align}
The lead electrons, the spinon, the holon (empty occupation) and the doublon (double occupation) are denoted by $c_{k\sig}$, $f_\sig$, $b$, and $d$, respectively. The antisymmetric tensor $\epsilon_{\sig\sig'}$ is defined as $\epsilon_{\uparrow\downarrow}=-\epsilon_{\downarrow\uparrow}=1$ and $\epsilon_{\uparrow\uparrow}=\epsilon_{\downarrow\downarrow}=0$. Double occupation is excluded for infinite $U$, thus in this case terms containing $d$ or $d^\dagger$ can be abandoned accordingly.

MIs can be modeled by the Heisenberg lattice $H_R=-J\sum_{i,j}(\frac{1}{2}S^+_iS^-_j+\frac{1}{2}S^-_iS^+_j+S^z_iS^z_j$), where $S^+_{i(j)} (S^-_{i(j)})$ is the raising (lowering) operator for the lattice spin, and $S^z_{i(j)}$ is the operator corresponding to the $z$-component of the spin. Using the Holstein-Primakoff transformation~\cite{Holstein} and assuming large lattice spin limit, one can approximate $S^+_{i(j)}\approx\sqrt{2S_0}a_{i(j)}$ and $S^-_{i(j)}\approx\sqrt{2S_0}a^\dagger_{i(j)}$. Here, $S_0$ is the length of the lattice spins, which are assumed to satisfy a condition $2S_0\gg \lan a^+_{i(j)} a_{i(j)} \ran$. It is clear from this formulation that the magnon operator $a^\dagger_{i(j)}$ ($a_{i(j)}$) adds (lessens) the $z$-component of the magnon by one unit. After a Fourier transform into the momentum space, the Hamiltonian can be approximated by the free Bose gas, $H_R\approx\sum_q\hbar\omega_q a^\dagger_q a_q$. Here we drop a constant from the term $\sum_{i,j}S^z_iS^z_j$.

Following Refs.~\cite{Ren,Takahashi,Adachi,Steven,Sothmann}, we describe the magnon-QD coupling by the \emph{s}-\emph{d} exchange interaction
\beq
H_{sd}=-\sum_{q}J_q \left[S_q^-c_{d\uparrow}^\dagger c_{d\downarrow}+S_q^+c_{d\downarrow}^\dagger c_{d\uparrow}\right],
\eeq
where $S_q^+=\sqrt{2S_0}a_q$, $S_q^-=\sqrt{2S_0}a_q^\dagger$ are the lattice spin operators in the momentum space, and $J_q$ denotes the effective exchange interaction. Here, we do not express the electron operators $c_{d\uparrow(\downarrow)}$ in QD in the slave-boson representation, since the original form makes the derivation of spin current more concise (See Appendix~\ref{ap:sc}). From the perspective of electrons in QD, the two terms represent emission and absorbtion of magnons respectively. The coupling with the $z$-component $-J_qS_q^z(c_{d\uparrow}^\dagger c_{d\uparrow}-c_{d\downarrow}^\dagger c_{d\downarrow})$ is absorbed into the energy level in QD, resulting in a difference in energy for the two spins.

The spin current in MI is carried by magnons and it can be defined as $I_S\equiv\frac{d}{dt}\lan\sum_q a^\dagger_q a_q\ran$. Since each magnon carries a unit of up spin, this definition implies that the reference direction of spin currents is defined as the up spin moving from metal to QD or the inverse for the down spin. To the lowest order of $J_q$, the spin current is given by (see Appendix~\ref{ap:sc})
\begin{align}
I_S=\frac{2S_0}{\pi\hbar}\int_{0}^{+\infty}d\omega_qJ_q^2\rho_R(\omega_q)\int_{-\infty}^{+\infty}d\omega
\left[(1+N_R(\omega_q))G_{d\downarrow}^<(\omega+\omega_q)G_{d\uparrow}^>(\omega)\right.\nonumber\\
-\left. N_R(\omega_q)G_{d\downarrow}^>(\omega+\omega_q)
G_{d\uparrow}^<(\omega)\right],\label{spincurr}
\end{align}
where $\rho_R(\omega_q)$ denotes the DOS of magnons and $N_R(\omega_q)=[e^{\hbar\omega_q/k_BT_R}-1]^{-1}$ is the Bose-Einstein distribution. $G_{d\sig}^>$ ($G_{d\sig}^<$) is the bigger (lesser) Green's function of electrons in QD. If the metal lead couples with MI directly, this expression is equivalent to that in Ref.~\cite{Ren}.

In the slave-boson representation, a term $i\lambda (Q-1)$ is added to the original Hamiltonian so that quantities can be calculated in the unconstrained ensemble. The physical quantities are obtained by integrating over $\lambda$ or otherwise thorough a projection procedure~\cite{Coleman}, which  enforces the constrain
\beq
Q=\sum_\sig f_\sig^\dagger f_\sig + d^\dagger d + b^\dagger b=1\label{constrain}.
\eeq
The expectation value of an operator $O$ is given by~\cite{Wingreen}
\begin{align}
\lan O\ran_{Q=1}&=\frac{Z_{Q=0}}{Z_{Q=1}}\frac{\beta}{2\pi}\int_{-\pi/\beta}^{\pi/\beta}e^{i\beta}\lan O\ran_{i\lambda}\nonumber\\
&=\frac{Z_{Q=0}}{Z_{Q=1}}\lan O\ran^{(1)}_{i\lambda},
\end{align}
where $\lan O\ran_{i\lambda}$ is the average taken over the unconstrained ensemble, and $\lan O\ran^{(1)}_{i\lambda}$ is the coefficient of the term of order $e^{-i\beta\lambda}$ in $\lan O\ran_{i\lambda}$. $Z_{Q=0/1}$ denotes the partition function in the subspace with constrain $Q=0/1$. The normalization can be obtain from the identity $\lan O \ran_{Q=1}=1$, which states
\begin{align}
\frac{Z_{Q=1}}{Z_{Q=0}}&=\lan b^\dagger b\ran_{i\lambda}^{(1)}+\lan d^\dagger d\ran_{i\lambda}^{(1)}
+\sum_\sig \lan f_\sig^\dagger f_\sig\ran_{i\lambda}^{(1)},\nonumber\\
&=\frac{i}{2\pi}\int_{-\infty}^{+\infty} d \omega [B^<(\omega)+D^<(\omega)-\sum_\sig G_{f\sig}^<(\omega)].
\end{align}
Here, $B$ and $D$ denote holon and doublon Green's functions.

By substituting the electron Green's functions in Eq.~\ref{spincurr} with the spinon functions $G^\gtrless_{f\sig}$ and multiplying the right-hand term with the normalization factor $Z_{Q=0}/Z_{Q=1}$, we obtain the spin current in the slave-boson representation as shown in Eq.~\ref{slavesc}. Technically, the holon and doublon are eliminated in the (anti)commutative relations and the constrain in Eq.~\ref{constrain}, as shown in Appendix~\ref{ap:sc}. Their absence can be understood intuitively because they do not carry spin.

The self-energy of electrons in the QD due to the interaction with magnons must involve an even number of spin flipping processes to conserve the spin, so its lowest-order approximation is of order $O(J_q^2)$. We may neglect this self-energy since we consider the situation that $J_q$ is much smaller than $V_k$. As a result, the NCA calculation of the Green's functions is similar to that of the Anderson impurity. For the case with an infinite-\emph{U}, the formulations and iteration procedures are the same as those presented in Refs.~\cite{Wingreen} and ~\cite{Hettler}, while for cases with finite correlations they are more complicated due to the presence of doublons.

Between the two equivalent iteration procedures, we use the one in Ref.~\cite{Hettler}. Since the lesser functions are of order $O(e^{-i\beta\lambda})$ and the retarded functions are $O(1)$, the former can be dropped from identities that relate the retarded functions with the bigger and lesser functions~\cite{Wingreen}. Hence, the usual identities are reduced to
\begin{align}
&\Sigma^>(\omega)=2i\mathrm{Im} \Sigma^R(\omega),\label{simple1}\\
&G^>(\omega)=2i\mathrm{Im} G^R(\omega),\label{simple2}
\end{align}
where $\Sigma$ and $G$ denote general self-energies and Green's functions. With the above relations, the expressions of the bigger self-energies (see section IV) are the relations between the imaginary component of the retarded self-energies and Green's functions. Together with the Kramers-Kronig relation and $G^R(\omega)=(\omega-\vep-\Sigma^R+i\eta)^{-1}$, those equations constitute the iteration for the bigger Green's functions. Given the retarded Green's functions, the lesser Green's functions can be obtained with an iteration made of $G^<(\omega)=\Sigma^<(\omega)|G^R(\omega)|^2$ and the expressions of the lesser self-energies.

\section {Extreme cases with $U=0$ and $U=\infty$}
\label{sec:nc}

To evaluate the effect of electronic correlations in QDs, we take two limiting cases into account, i.e., $U=0$ and $U=\infty$. When the Coulomb correlation is absent, the Green's functions can be calculated exactly. Assuming the coupling between the QD and metal lead in Lorentzian is spin independent, the bigger and lesser self-energies are given by
\begin{align}
&\Sigma_\sig^>(\omega)=-i\Gamma(\omega)(1-f_\sig(\omega)),\\
&\Sigma_\sig^<(\omega)=i\Gamma(\omega)f_\sig(\omega),
\end{align}
with
\beq
\Gamma({\omega})=2\pi\sum_{k} |V_k|^2 \delta(\omega-\vep_k)\equiv\Gamma_0\frac{W_L^2}{(\omega-\vep_L)^2
+W_L^2}.
\eeq
Here, $f(\omega)$ is the Fermi-Dirac distribution. We take $\Gamma_0=1.0$ eV, $\vep_L=0$, and $W_L=10$ eV throughout this work. Furthermore, we set the Fermi level $E_F=2.0$ eV for both spins in the metal lead, and neglect the difference in the chemical potential caused by the cumulation of spins~\cite{Valet,Maekawa}. As a result, the retarded self-energy takes the form
\beq
\Sigma_\sig^R=\frac{\Gamma(\omega)}{2W_L}(\omega-\vep_L-iW_L),
\eeq
and finally $G^\gtrless(\omega)=\Sigma^\gtrless(\omega)|G^R(\omega)|^2$ gives the lesser and bigger Green's functions.

For the insulating magnetic lead, we follow the Refs.~\cite{Ren,Adachi,Tupitsyn,Kreisel} and adopt an Ohmic spectrum, $J_q^2\rho_R(\omega_q)=\alpha \frac{\omega_q}{\omega_c}e^{-\omega_q/\omega_c}$, with $\alpha=10$ and $\omega_c=50$ meV. Since here only $J_q^2\rho_R(\omega_q)$ is fixed, there is a freedom to satisfy the weak coupling assumption by adopting a small $J_{q}$. The length of the lattice spins is set as $S_0=16$. The difference in the QD level resulting from $\sum_q -J_qS_q^z(c_{d\uparrow}^\dagger c_{d\uparrow}-c_{d\downarrow}^\dagger c_{d\downarrow})$ is taken to be $\vep_\downarrow-\vep_\uparrow=0.1$ eV, which is much smaller than $\Gamma_0$. When the correlation is strong, the difference has little impact on the main results. We confirm this point by setting $\vep_\downarrow=\vep_\uparrow$ at different values and find that only small quantitative changes are brought about. For infinite-\emph{U} Coulomb correlation, the formulations and iteration procedure can be referred to Refs.~\cite{Wingreen,Hettler}.

\begin{figure}
\centering
\includegraphics[width=0.7\textwidth]{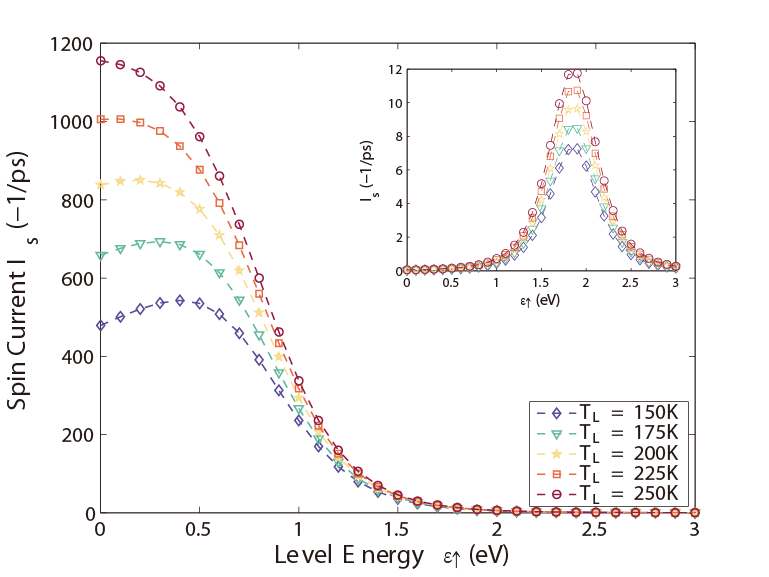}\\
\caption{Dependence of the spin-Seebeck current $I_S$ in the infinite-$U$ QD with the constant temperature difference $\Delta T=50$ K on the QD level energy for various $T_L$. In the inset, the results for the non-correlated QD are plotted for comparison.}\label{Figure_2}
\end{figure}

To see how the shift of the QD level affects the spin-Seebeck current, we calculated their relations versus $\vep_\uparrow$ for some values of $T_L$ and a fixed temperature gradient $\Delta T$ $(=T_{R}-T_{L})$ = 50 K, and the results are illustrated in Fig.~\ref{Figure_2}. Results for QD with $U=0$ are also included in the inset for comparison. At the first glance, we find that the magnitude of the spin current in the infinite-$U$ QD is larger, by two orders in magnitude, than that in the non-correlated QD. Although the hight of the current peak increases $T_{L}$ for both $U=0$ and $U=\infty$, their positions show different variation trends. For the infinite-$U$ QD, the peaks locate well below the Fermi level ($E_F=2.0$ eV) in the metal lead, and moves lower with the increase of $T_{L}$. As for the non-correlated QD, the maxima all occur at $\vep_\uparrow = 1.9$ eV regardless of the temperature variation, meaning that the related QD level for spin-down electrons $\vep_\downarrow$ just locates at the Fermi level of the metal lead.

\begin{figure}
\centering
\includegraphics[width=0.65\textwidth]{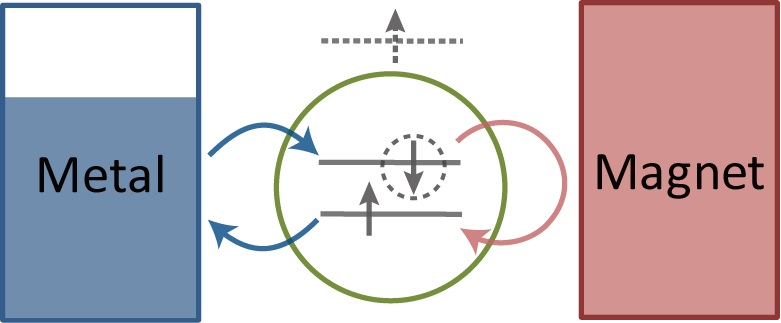}\\
\caption{Schematic of how the magnetic insulator drive a spin current in the metal lead. Three processes take place in turn: 1) A spin-down electron tunnels into the QD; 2) the spin is flipped by absorbing a magnon; 3) the spin-up electron tunnels back to the metal.}\label{Figure_3}
\end{figure}

To give a physical picture for the understanding of the large difference in the magnitudes of $I_S$ for these two extreme cases, we first explore the physical mechanism to generate the relatively small spin current in the case with $U=0$. The detailed process to drive a spin-up current from MI to the metal lead is illustrated in Fig.~\ref{Figure_3}. First, due to the coupling with the MI having upward polarization, the level of the spin-down state is a little higher, indicating that the spin-up state is more likely to be occupied regardless of the absolute positions of the two levels. When a spin-down electron tunnels into QD, and after absorbing a magnon from MI, the electron flips its spin direction and then tunnels back to the metal lead. Of course, the inverse process occurs at the same time. Nevertheless, constricted by the polarization direction of MI and the temperature biased with $T_{R}>T_{L}$, the inverse process is largely suppressed. As a result, a unit of net up spin is added to the metal and a spin current is produced. Obviously, the spin flipping in QD is a key step to generate the spin current. According to the exclusion principle, however, the spin flipping in the present device may occur only when the spin-down state is occupied while the spin-up state is empty (we call this case singlet occupation below). As for the current QD-based device with $U=0$, when both levels in QD are higher (lower) than the Fermi level in the metal lead, the empty (double) occupation in the QD levels will occur, which suppresses the spin flipping in QD and decreases the spin current. For $\vep_\uparrow = 1.9$ eV, the spin-down state in QD just locates at the Fermi level in the metal lead, and the singlet occupation in QD reaches its maximum probability. As a result, the spin flipping is largely enhanced and the maxima in spin current are achieved.


From the above discussion, we well know that when both levels in QD are below the Fermi level in the metal lead, double occupation occurs and spin flipping is suppressed. However, as a Coulomb interaction with $U>E_F-\vep_\uparrow$ is introduced, the situation is much different. For the case where a spin-up electron already exists in the QD, the level of the spin-down state shifts upwards by the Coulomb interaction, resulting in the appearance of a single occupation of the spin-up state in QD. Nevertheless, restricted by the upward polarization in MI, it is very difficult for the spin-up electron in QD to absorb a magnon from MI and to flip its spin direction. However, for another case when a spin-down electron already locates in the QD, the energy level of the spin-up state is lifted by the Coulomb interaction to $\vep_{\uparrow}+U$ and it is unlikely to be occupied. As a sequence, the single occupation of the spin-down state occurs in QD and the spin current is generated again. It should be noted that for $\vep_{\uparrow(\downarrow)}>E_F$, there is only empty occupation occurring in QD for any value of $U$ and the spin current is suppressed. As $U$ is large enough, the single occupation in QD can be achieved by setting the levels in QD well below the Fermi level, which is confirmed by our numerical result that all the current peaks in the case with $U=\infty$ locate below the Fermi level as shown in Fig.~\ref{Figure_3}. Besides, although $\vep_{\downarrow}>\vep_{\uparrow}$ implies that the level of the spin-up state tends to be occupied, the single occupation of the spin-down state, which contributes to the generation of spin current, remains as highly probable since the difference $\vep_{\downarrow}-\vep_{\uparrow}$ is small.

Keeping in mind that the singlet occupation of the spin-down state is a prerequisite for
the formation of spin current, the significant enhancement in spin current by the Coulomb
interaction can be easily understood as follows. In the case with $U=0$, due to $\vep_{\downarrow}>\vep_{\uparrow}$ and the fact that the Fermi-Dirac distribution declines sharply around the Fermi level with the increase of energy, the singlet-occupation probability of the spin-down state is lower than that of the spin-up state. In the case with $U=\infty$, as discussed above, due to the
Coulomb interaction, the empty and double occupations in QD are largely excluded, and the singlet-occupation probability of the spin-down state can be compared with that of the spin-down state. As a result, the spin current in the case with $U=\infty$ is remarkably enhanced compared to that in the case with $U=0$.

Beside the view point of the dynamical process described above, the transport properties are also reflected in the DOS of QD electrons. Intuitively, only the electronic states close to the Fermi level conduct the spin current, since spin flipping requires that the energy level of QD is occupied by one electron. To make this clear, we set $T_{L}=0$ K and then Eq.~\ref{spincurr} is reduced to
\beq
I_S=-\frac{8\pi S_0}{\hbar}\int_{0}^{+\infty}d\omega_qJ_q^2\rho_R(\omega_q)N_R(\omega_q)\int_{\mu-\omega_q}^{\mu}d\omega \rho_{d\downarrow}(\omega+\omega_q)
\rho_{d\uparrow}(\omega),\label{zeroT}
\eeq
where $\rho_{d\uparrow(\downarrow)}$ is the DOS of the electrons in QD. From the above expression, we can find that the spin current is contributed by the interfacial exchange interaction strength $J_q$, the number of magnons $N_R$ in MI, and the DOS in QD. Since the Ohmic spectrum function decays rapidly for $\omega_q>\omega_c$, and the same parameter values are applied in the non-correlated and correlated cases, the amplification of spin current is mostly ascribed to the ingredient of the electronic states in the range of $({E_F}-\omega_c, {E_F}+\omega_c)$. It is noted that this simplification is based on the fact that the Fermi-Dirac distribution reduces to the unit step function at zero temperature. Owing to the sharp descending feature of the Fermi-Dirac function, this observation should hold for reasonable non-zero temperatures.

To shed light on the temperature independence of the peak positions in the spin current for the non-corrected QD, we refer to a simplified Eq.~(\ref{spincurr}). Since the narrow range contributes for the most part to the integral in the calculation of spin current, and the spectral function is smooth in the non-correlated QD, the temperature independent part of the integrand in Eq.~(\ref{spincurr}) can be approximated by a function of $\vep_{\uparrow}$ that stays constant versus $\omega$, i.e., $C_L(\vep_{\uparrow})\approx \rho(\omega+\omega_q)_{d\downarrow}\rho(\omega)_{d\uparrow}$. Putting $C_{L}$ out of the integral expression and using the equality $\int d\omega f_L(\omega+\omega_q)\left[1-f_L(\omega)\right]=N_{L}(\omega_q)$, we can cast the temperature dependent part into a function
\beq
F(T_L, T_R)=\int_{0}^{+\infty}d\omega_qJ_q^2\rho_R(\omega_q)\omega_q\left[ N_{L}(\omega_q)-N_{R}(\omega_q)\right],
\eeq
which does not depend on $\vep_{\uparrow}$. With these reasonable simplifications, we find that the positions of the maxima in the spin current are determined by $C_{L}(\vep_{\uparrow})$ and are thus independent of temperature.

\begin{figure}
\centering
\includegraphics[width=0.8\textwidth]{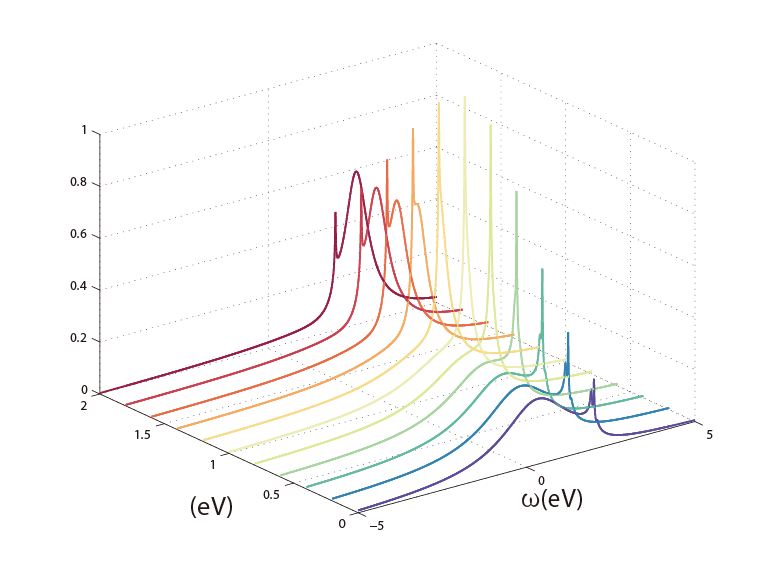}\\
\caption{Two peaks of the DOS of the spin-up states: a broad peak around the QD level and a narrow one at the Fermi level. The temperature of the metal lead is set at $T_{L}=200$ K.}\label{Figure_4}
\end{figure}

As a correlated impurity coupled with metal leads, the DOS in QD is composed of three peaks: a broadening peak around the level in QD, i.e, $\omega=\vep$, a narrow peak at the Fermi level of the lead, which can be referred to as the Kondo peak, and the peak around $\omega=\vep+U$~\cite{Wingreen}. For a QD with $U=\infty$, the third peak lies far away from the effective energy region and can thus be neglected. Since the DOSs for up- and down-spin are similar, we calculate the spin-up DOS ($\rho_{\uparrow}$) and the results are shown in Fig.~\ref{Figure_4}. According to Eq.~(\ref{zeroT}), the electronic states near the Fermi level provide the major channels to conduct the spin current. Here, the developed Kondo peak brings an alternative perspective on why the spin current is significantly amplified further by the Coulomb interaction, besides an intuitive understanding based on the driving process plotted in Fig.~\ref{Figure_3}.

The magnitude of the spin current is determined by both DOS and the occupation of electronic states in QD near the Fermi level of the metal lead, where the Fermi-Dirac distribution varies dramatically. As a result, the magnitude depends on the details of the DOS for both spins, and no simple correspondence between the current magnitude and the height of the Kondo peak can be drawn. Still, the variation of DOS with the level of QD provides a clue as to how a magnitude maximum is reached. When the level in QD is deeply below the Fermi level of the metal lead, the two peaks are separated. As the level in QD shifts toward the Fermi level, the two peaks move closer and result in a higher DOS around the Fermi level. Thus, a maximum of the spin current is achieved. Note that the DOS for a non-degenerated QD in the slave-boson representation is not quantitatively reliable because the vertex correlation is neglected~\cite{Wingreen}. Nevertheless, the calculation of spin current is not much affected by this neglect, since there is no interaction between spinons and bosons according to Eqs.~(\ref{spinon}) and ~(\ref{slavesc}).

\section{Finite Correlations in the QD}

\label{sec:fc}
In order to study the influence of finite correlations and mutually corroborate the results for the above two extreme cases, we now generalize the NCA to the finite-$U$ case. As is known, the NCA is defined by the first-order self-energies, which are represented diagrammatically in Fig.~\ref{Figure_5}. As presented in Appendix~\ref{ap:se}, their bigger and lesser components can be derived through the equation of motion~\cite{Stefanucci}, namely,
\begin{align}
&\Sigma_{f\sig}^\lessgtr(\omega)=\sum_k\frac{i|V_k|^2}{2\pi}\left[\int_{-\infty}^{+\infty} d\omega' g_{k\sig}^\lessgtr(\omega-\omega')B^\lessgtr(\omega')-\int_{-\infty}^{+\infty} d\omega' g_{k\sig'}^\gtrless(\omega'-\omega)D^\lessgtr(\omega')\right],\label{self1}\\
&\Pi^\lessgtr(\omega)=-\frac{i}{2\pi}\sum_{k,\sig}|V_k|^2\int_{-\infty}^{+\infty} d\omega' g_{k\sig}^\gtrless(\omega'-\omega)G_{f\sig}^\lessgtr(\omega'),\label{self2}\\
&\Lambda^\lessgtr(\omega)=\frac{i}{2\pi}\sum_{k,\sig\neq\sig'}|V_k|^2\int_{-\infty}^{+\infty} d\omega' g_{k\sig}^\lessgtr(\omega-\omega')G_{f\sig'}^\lessgtr(\omega').
\end{align}
When the second term in Eq.~(\ref{self1}) is dropped, Eqs.~(\ref{self1}) and ~(\ref{self2}) are the same as the formulations given in Ref.~\cite{Wingreen}. Following the iteration procedure discussed in Section II, the lesser and bigger Green's functions can be obtained. The accuracy is checked by unity of the spectral functions, which is satisfied to better than $0.1\%$.

\begin{figure}
\centering
\includegraphics[width=0.6\textwidth]{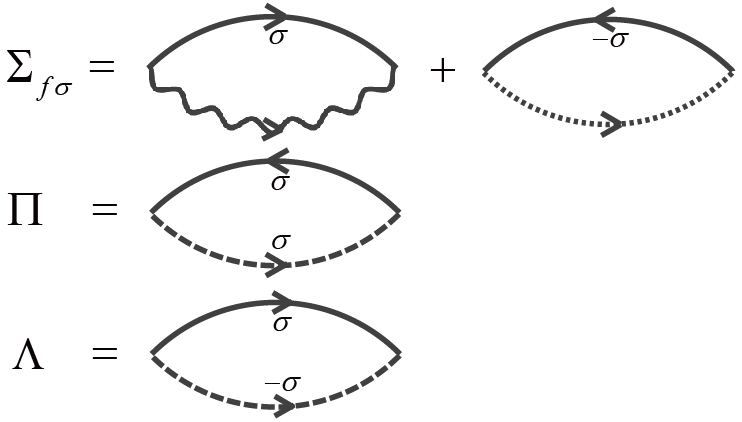}\\
\caption{Diagrammatic representation of the NCA self-energies. $\Sigma_{f\sig}$ is the spinon self-energy. The solid line, wavy line, and dotted line represent propagators of the lead electrons, holon and doublon respectively. $\Pi$ and $\Lambda$ are the holon and the doublon self-energies, where the dashed line denotes the spinon propagator.}\label{Figure_5}
\end{figure}

\begin{figure}
\centering
\includegraphics[width=0.7\textwidth]{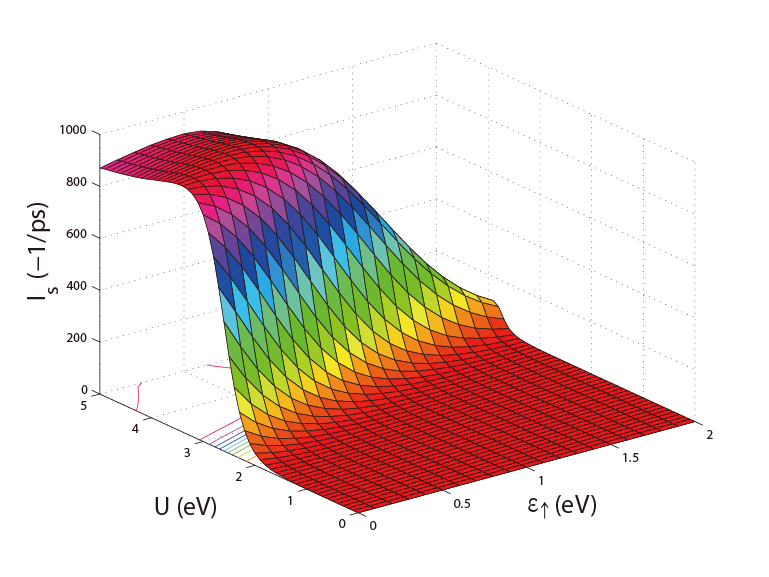}\\
\caption{Variation of the spin-Seebeck current $I_S$ vs the QD level $\vep_{\uparrow}$ and correlation $U$. Temperature of the metal lead $T_{L}$ and the temperature gradient $\Delta T$ are fixed at $200$ and $50$ K, respectively.}\label{Figure_6}
\end{figure}

In Fig.~\ref{Figure_6}, we show how the spin-Seebeck current is tuned by changing the energy level and the Coulomb interaction in QD. For weak Coulomb interactions, the spin current is suppressed and shows a nice linear relationship with $\vep_{\uparrow}$. As the Coulomb correlation increases, the amplifying effect in the spin-Seebeck current starts to take place. When the correlation increases further, the situation becomes similar to that of infinite-$U$ QD. For a fixed correlation, the spin current increases till the energy level $\vep_{\uparrow}$ in QD reaches a critical value and then decreases when the level is further lowered. For a fixed level in QD and varying Coulomb correlation, the spin-Seebeck current also follows a similar pattern that first goes up and then down. As a result, the magnitude of the spin current develops a peak at $U\approx3.8$ eV and $\vep_\uparrow\approx0.4$ eV. Thus in the $U-\vep_{\uparrow}$ plane, there exist optimized structure parameters to achieve the maximum spin-Seebeck current in the current QD-based junction.

\begin{figure}
\centering
\includegraphics[width=0.7\textwidth]{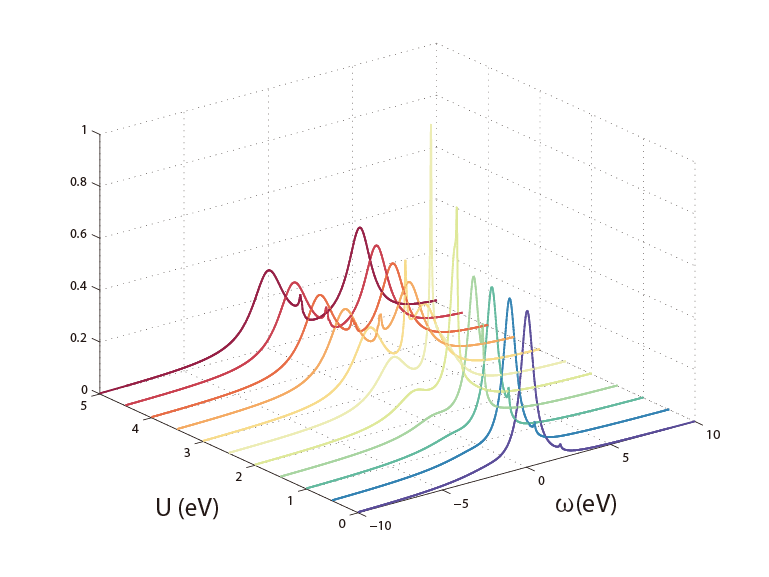}\\
\caption{DOS of spin-up electrons for some correlations ranging from $U=0\sim5$ eV. The QD level is fixed at $\vep_{\uparrow}=0$ eV and temperature of the metal lead is $T_{L}=200$ K.}\label{Figure_7}
\end{figure}

\begin{figure}
\centering
\includegraphics[width=1.0\textwidth]{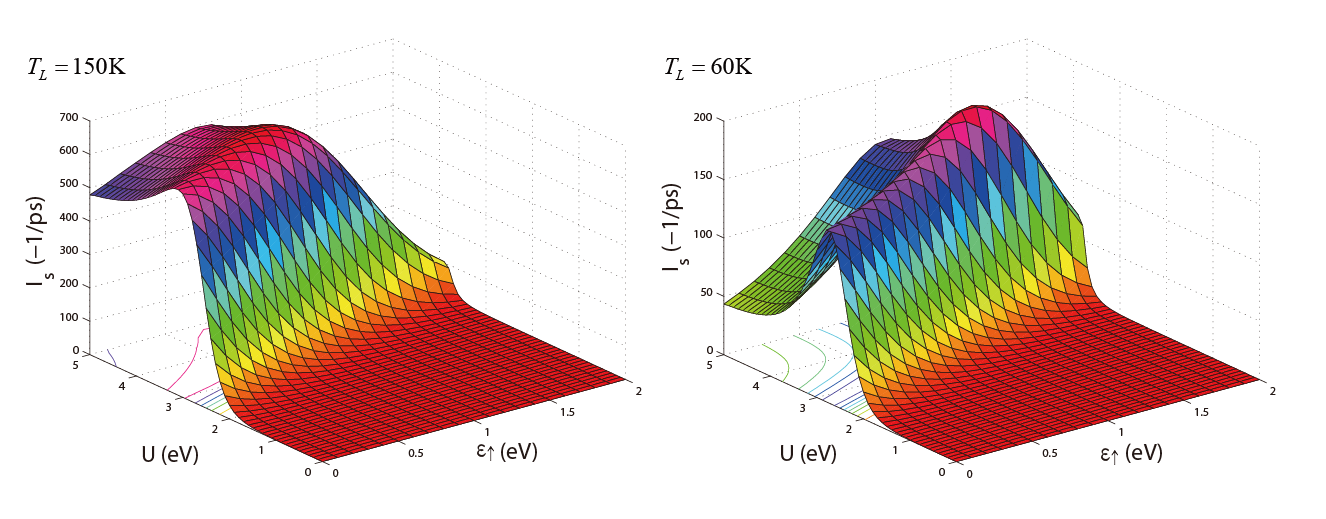}\\
\caption{Variation of the position of the spin current peak vs temperature of the metal lead. Results for temperatures $T_{L}=150, 60$ K are presented with temperature gradient $\Delta T=50$~K. The corresponding QD level is at $\vep_{\uparrow}\approx0.4, 0.8$ eV, respectively. In both cases the peak is located at $U\approx3.8$ eV, the same as it is in Fig.~\ref{Figure_6}.}\label{Figure_8}
\end{figure}

The emergence of a maximum in the spin-Seebeck current can be understood from the influence of the Coulomb interaction on the DOS near the Fermi level of the metal lead. In the present QD-based junction with a finite Coulomb correlation, the DOS is also calculated and plotted in Fig.~\ref{Figure_7}, in which three different types of DOS peaks clearly appear. For a weak Coulomb interaction, the DOS is approximately the ordinary level broadening around $\vep_{\uparrow}=0$. When $U$ increases, besides that the narrow peak at the Fermi level ($E_F=$ 2.0eV) is steepened, the rise of the peak around $\vep_{\uparrow}+U$ results in a high intensity of the DOS at the Fermi level. With a further increase of $U$, intensification around $\vep_{\uparrow}+U$ dilutes the DOS to other regions. As a result, the maximum height of the DOS peak is achieved when the correlation is at a critical value. Comparing with Figs.~\ref{Figure_6} and ~\ref{Figure_7}, we find that although there is no precise correspondence between the magnitude of the spin current and the height of the DOS peak at the Fermi level, they indeed corresponds with each other in the variation trend.

Finally, to see how the positions of the maximum values of the spin-Seebeck current vary with the change of temperature in the metal lead, we present the calculated results about the variation of spin current versus the Coulomb interaction $U$ and the energy level $\vep_{\uparrow}$ in QD. To examine the trend at low temperatures, we purposefully study the case with $T_{L}$ close to the Kondo temperature, which can be estimated from a degenerate QD, i.e., $T_k\approx W_L[\Gamma_0/2\pi(\mu-\vep_0)]^{1/2}\mathrm{exp}[-\pi(\mu-\vep_0)/\Gamma_0]$ with $\vep_0$ the dot level~\cite{Wingreen}. In our device parameter settings, the Kondo temperature is about $61$ K at $\vep_0=0$ and increases when the level in the QD shifts up. According to Fig.~\ref{Figure_6} and~\ref{Figure_8}, we find that the lower temperature corresponds to a higher level in the QD (i.e., bigger $\vep_{\uparrow}$), which is consistent with the variation trends shown in Fig.~\ref{Figure_2}, while the location of $U$ is insensitive to the temperature variation and is located at $U=3.8$ eV in all cases. It is believed that there is an optimized value of $U$ that can maximize the tunability of the spin-Seebeck current.

\section{Summary}

Through structuring a spin caloritronic device based on a correlated QD coupled to a metal lead and a MI, we have studied the spin-Seebeck current through the QD junction in different situations, including non-correlated QD, infinite Coulomb-correlation QD, and finite Coulomb-correlation QD. The results show that the spin-Seebeck current in the infinite Coulomb interacted QD is remarkably larger than in the non-correlated case. Moreover, as the energy level of the QD locates in a range below the Fermi level of the metal lead, the spin-Seebeck current can be tuned in a wide range, and this process can be realized easily by adjusting the gate voltage in experiments. Thus the proposed spin-Seebeck device can work as a thermovoltaic transistor. Besides, we find that there are optimal correlation strengths and energy levels of the QD which can maximize the spin-Seebeck current. These results put forward a direction for the design of a tunable spin caloritronic device without changing the temperature gradient.

\begin{acknowledgments}
This work is supported by the National Natural Science Foundation of China (Nos. 11274128, 10804034 and 11074081). Work at UCI was supported as part of the SHINES, an Energy Frontier Research Center funded by the U.S. Department of Energy, Office of Science, Basic Energy Sciences under Award SC0012670.
\end{acknowledgments}

\appendix
\section{Derivation of the spin-Seebeck current}
\label{ap:sc}

In the following derivations, we set $\hbar=1$ and make it explicit in the last when it is necessary. For notational simplicity, the notation of contour-time-order is not denoted explicitly. The averages should be understood as contour-time-ordered.

The spin current is defined as
\begin{align}
I_S&=\frac{d}{dt}\lan \sum_q a_q^\dagger a_q\ran\nonumber\\
&=i\lan [H,\sum_q a_q^\dagger a_q]\ran\nonumber\\
&=i\sum_q \sqrt{2S_0}J_q \left(\lan a^\dagger_q c_{d\uparrow}^\dagger c_{d\downarrow}\ran-\lan a_q c_{d\downarrow}^\dagger c_{d\uparrow}\ran\right).
\end{align}
Since the two terms are the complex conjugate of each other, only one is needed and we evaluate the second term. The average of an operator $O(t)$ can be taken on either the forward $\lan O(t^+)\ran$ or the backward $\lan O(t^-)\ran$ time branch, or half-and-half $(\lan O(t^+)\ran+\lan O(t^-)\ran)/2$. We adopt the last one, so to the first order in $J_q$,
\begin{align}
&\lan a_q(t) c_{d\downarrow}^\dagger(t) c_{d\uparrow}(t)\ran\nonumber\\
=&i \sqrt{2S_0}J_q \int_{kc} dt'\frac{1}{2}(C^T(t,t')+C^{\tilde{T}}(t,t')+C^>(t,t')+C^<(t,t'))\nonumber\\
=&i \sqrt{2S_0}J_q \int_{kc} dt'(C^>(t,t')+C^<(t,t'))\nonumber\\
=&i \sqrt{2S_0}J_q \int_{-\infty}^{+\infty} dt'(C^>(t,t')-C^<(t,t')).
\end{align}
where
\beq
C(t,t')=\lan a_q(t) c_{d\downarrow}^\dagger(t) c_{d\uparrow}(t) a_q^\dagger(t') c_{d\uparrow}^\dagger(t') c_{d\downarrow}(t')\ran,\label{composite}
\eeq
and $kc$ denotes the Keldysh contour. The minus in the last step arises due to reversing of the integral range, since the default range of $t'$ is $(+\infty,-\infty)$ when it is on the backward branch. In steady states, the Wick theorem gives
\begin{align}
\lan a_q(t) c_{d\downarrow}^\dagger(t) c_{d\uparrow}(t)\ran
=-\sqrt{2S_0}J_q \int_{-\infty}^{+\infty} dt'\left[A_q^>(t,t')G_{d\downarrow}^<(t',t)G_{d\uparrow}^>(t,t')\right.\nonumber\\
\left.-A_q^<(t,t')G_{d\downarrow}^>(t',t)G_{d\uparrow}^<(t,t')\right].
\end{align}
Here, $A_q$ is the magnon Green's function, whose bigger and lesser components are given by
\begin{align}
A_q^>(t,t')&=-i(1+N_R(\omega_q))e^{-i\omega_q(t-t')},\\
A_q^<(t,t')&=-iN_R(\omega_q)e^{-i\omega_q(t-t')},
\end{align}
where $N_R(\omega_q)$ denotes the Bose-Einstein distribution for the magnon. Fourier transforms of the electron Green's functions lead to
\begin{align}
&\lan a_q(t) c_{d\downarrow}^\dagger(t) c_{d\uparrow}(t)\ran\nonumber\\
=&\frac{i\sqrt{2S_0}J_q}{(2\pi)^2}\int_{-\infty}^{+\infty} dt'd\omega d\omega'(1+N_R(\omega_q))e^{-i\omega_q(t-t')}G_{d\downarrow}^<(\omega)e^{-i\omega(t'-t)}G_{d\uparrow}^>(\omega')
e^{-i\omega'(t-t')}\nonumber\\
&-\frac{i\sqrt{2S_0}J_q}{(2\pi)^2}\int_{-\infty}^{+\infty} dt'd\omega d\omega'N_R(\omega_q)e^{-i\omega_q(t-t')}G_{d\downarrow}^>(\omega)e^{-i\omega(t'-t)}
G_{d\uparrow}^<(\omega')e^{-i\omega'(t-t')},
\end{align}
In the steady state, using the Markov condition $\lim_{t\to+\infty}e^{i\omega t}=1$, we get
\begin{align}
\lan a_q c_{d\downarrow}^\dagger c_{d\uparrow}\ran
=\frac{i\sqrt{2S_0}J_q}{2\pi}\delta(\omega_q-\omega+\omega')\iint d\omega d\omega'\left[(1+N_R(\omega_q))G_{d\downarrow}^<(\omega)G_{d\uparrow}^>(\omega')\right.\nonumber\\
\left.-N_R(\omega_q)G_{d\downarrow}^>(\omega)
G_{d\uparrow}^<(\omega')\right].
\end{align}
Without referring to the Markov condition, one can obtain the same result by Fourier transforming $\lan a_q c_{d\downarrow}^\dagger c_{d\uparrow}\ran(t)$ and performing the integral of $\lan a_q c_{d\downarrow}^\dagger c_{d\uparrow}\ran(\omega)$ over the frequencies. Representing the sum $\sum_q$ as an integral $\int_{0}^{+\infty}\rho_R(\omega_q)d\omega_q$, we arrive at the expression
\begin{align}
I_S=\frac{2S_0}{\pi\hbar}\int_{0}^{+\infty}d\omega_qJ_q^2\rho_R(\omega_q)\int_{-\infty}^{+\infty}d\omega
\left[(1+N_R(\omega_q))G_{d\downarrow}^<(\omega+\omega_q)G_{d\uparrow}^>(\omega)\right.\nonumber\\
-\left. N_R(\omega_q)G_{d\downarrow}^>(\omega+\omega_q)
G_{d\uparrow}^<(\omega)\right].
\end{align}

In the slave boson representation, substituting $c_{d\sig}^\dagger=f_\sig^\dagger b+\epsilon_{\sig\sig'} f_{\sig'}d^\dagger$ into Eq.~(\ref{composite}), we have
\begin{align}
C(t,t')=&\lan a_q(t) a_q^\dagger(t') [f_\downarrow^\dagger(t) f_\uparrow(t) b(t) b^\dagger(t)-f_\uparrow(t) f_\downarrow^\dagger(t) d^\dagger(t) d(t)]\nonumber\\
&[f_\uparrow^\dagger(t') f_\downarrow(t') b(t') b^\dagger(t')-f_\downarrow(t') f_\uparrow^\dagger(t') d^\dagger(t') d(t')]\ran\nonumber\\
=&\lan a_q(t) a_q^\dagger(t')f_\downarrow^\dagger(t) f_\uparrow(t)[1+b^\dagger(t)b(t)+d^\dagger(t) d(t)]\nonumber\\
&f_\uparrow^\dagger(t') f_\downarrow(t')[1+b^\dagger(t')b(t')+d^\dagger(t') d(t')]\ran\nonumber\\
=&\lan a_q(t) a_q^\dagger(t')f_\downarrow^\dagger(t) f_\uparrow(t)[2-f_\uparrow^\dagger(t) f_\uparrow(t)-f_\downarrow^\dagger(t) f_\downarrow(t)]\nonumber\\
&f_\uparrow^\dagger(t') f_\downarrow(t')[2-f_\uparrow^\dagger(t') f_\uparrow(t')-f_\downarrow^\dagger(t') f_\downarrow(t')]\ran\nonumber\\
=&\lan a_q(t) a_q^\dagger(t')f_\downarrow^\dagger(t) f_\uparrow(t)
f_\uparrow^\dagger(t') f_\downarrow(t')\ran.\label{spinon}
\end{align}
Accordingly, the spin current is given by
\begin{align}
I_S=\frac{Z_{Q=0}}{Z_{Q=1}}\frac{2S_0}{\pi\hbar}\int_{0}^{+\infty}d\omega_qJ_q^2\rho_R(\omega_q)
\int_{-\infty}^{+\infty}d\omega
\left[(1+N_R(\omega_q))G_{f\downarrow}^<(\omega+\omega_q)G_{f\uparrow}^>(\omega)\right.\nonumber\\
-\left. N_R(\omega_q)G_{f\downarrow}^>(\omega+\omega_q)
G_{f\uparrow}^<(\omega)\right].\label{slavesc}
\end{align}

\section{Derivation of self-energies}
\label{ap:se}
From the equation of motion of the Green's functions, one can read off the self-energies. For the spinon Green's function $G_{f\sig}(t,t')=-i\lan f_\sig(t)f^\dagger_\sig(t')\ran$,
\beq
i\frac{d}{dt}G_{f\sig}(t,t')=\theta(t,t')+\vep_\sig G_{f\sig}(t,t')-\sum_k i V^*_k\left[\lan c_{k\sig}(t)b(t)f^\dagger_\sig(t')\ran-\lan c^\dagger_{k\sig'}(t)d(t)f^\dagger_\sig(t')\ran\right].
\eeq
To the order $|V_k|^2$,
\begin{align}
&- i V^*_k\lan c_{k\sig}(t)b(t)f^\dagger_\sig(t')\ran\nonumber\\
=&-|V_k|^2\int_{kc}dt'' \lan c_{k\sig}^\dagger(t'')f_\sig(t'')b^\dagger (t'')c_{k\sig}(t)b(t)f^\dagger_\sig(t')\ran\nonumber\\
=&i|V_k|^2 \int_{kc}dt'' g_{k\sig}(t,t'')B(t,t'')G_{f\sig}(t'',t').
\end{align}
Here, $g_{k\sig}$ is the Green's function of the lead electrons and B the holon Green's function. Similarly,
\beq
i V_k\lan c^\dagger_{k\sig'}(t)d(t)f^\dagger_\sig(t')\ran\\
=-i|V_k|^2\int_{kc}dt'' g_{k\sig'}(t'',t)D(t,t'')G_{f\sig}(t'',t'),
\eeq
where $D$ is the doublon Green's function. Now we can read off the lesser and bigger self-energies for the spinon
\beq
\Sigma_{f\sig}^\lessgtr(t,t'')=\sum_k i|V_k|^2\left[g_{k\sig}^\lessgtr(t,t'')B^\lessgtr(t,t'')
+g_{k\sig'}^\gtrless(t'',t)D^\lessgtr(t,t'')\right].
\eeq
Their Fourier transforms read
\beq
\Sigma_{f\sig}^\lessgtr(\omega)=\sum_k\frac{i|V_k|^2}{2\pi}\left[\int_{-\infty}^{+\infty} d\omega' g_{k\sig}^\lessgtr(\omega-\omega')B^\lessgtr(\omega')-\int_{-\infty}^{+\infty} d\omega' g_{k\sig'}^\gtrless(\omega'-\omega)D^\lessgtr(\omega')\right].
\eeq
In the same manner, we can obtain the doublon and the holon self-energies.

\end{document}